\documentclass[useAMS,usenatbib]{mn2e}
\usepackage{graphicx}
\usepackage{eufrak}
\usepackage{eucal}
\usepackage{txfonts}

\begin{document}

\title[On the dust abundance gradients in late-type galaxies I]{On the dust abundance gradients in late-type galaxies:\\
I. Effects of destruction and growth of dust in the interstellar medium}

\author[Mattsson, Andersen \& Munkhammar]{Lars Mattsson$^1$\thanks{E-mail: mattsson@dark-cosmology.dk}, Anja C. Andersen$^1$ \& Joakim D. Munkhammar$^2$\\
$^1$ Dark Cosmology Centre, Niels Bohr Institute, University of Copenhagen, Juliane Maries Vej 30, DK-2100, Copenhagen \O, Denmark\\
$^2$ Dept. of Engineering Sciences, Solid State Physics, Uppsala University, Box 534, S-751 21 Uppsala, Sweden}

\date{}

\pagerange{\pageref{firstpage}--\pageref{lastpage}} \pubyear{2011}

\maketitle

\label{firstpage}

\begin{abstract}
We present basic theoretical constraints on the effects of destruction by supernovae (SNe) and growth of dust grains in the interstellar medium (ISM) on the radial distribution of dust in late-type
galaxies. The radial gradient of the dust-to-metals ratio is shown to be essentially flat (zero) if interstellar dust is not destroyed by SN shock waves and all dust is produced in stars. If there is net dust 
destruction by SN shock waves, the dust-to-metals gradient is flatter than or equal to the metallicity gradient (assuming the gradients have the same sign). Similarly, if there is net dust growth
in the ISM, then the dust-to-metals gradient is steeper than or equal to the metallicity gradient. The latter result implies that if dust gradients are steeper than metallicity 
gradients, i.e., the dust-to-metals gradients are not flat, then it is unlikely dust destruction by SN shock waves is an efficient process, while dust growth must be a significant mechanism for dust production.
Moreover, we conclude that {\it dust-to-metals gradients can be used as a diagnostic for interstellar dust growth in galaxy discs, where a negative slope indicates dust growth}.
\end{abstract}

\begin{keywords}
Galaxies: evolution, ISM; ISM: clouds, dust, extinction, evolution, supernova remnants;
\end{keywords}

\section{Introduction}
The lifetime of dust grains in the interstellar medium (ISM) is a critical parameter for the evolution of the dust component in a galaxy.
Shock-waves originating from supernovae (SNe) arguably contain enough energy to {destroy (or at least shatter)} dust grains as these waves propagate through 
the ISM. The time scale for such dust destruction {depends on several physical conditions, where} the supernova rate (SNR) and {efficiency of}
dust destruction in a  SN-shock {are the most important} \citep{McKee89, Draine90}. Shock destruction of dust grains has been considered quite 
efficient in many studies, e.g., \citet{Jones96,Jones04,Serra08}, suggest a grain lifetime of a few times $10^{8}$ yr {for many dust species, but note 
that a recent re-evaluation of dust lifetimes by \citet{Jones11} showed that the lifetimes of silicate grains may be comparable to the injection time scale 
of such grains}.

While this high dust-destruction efficiency seems consistent with the Milky Way (solar neighbourhood), it has been shown by several authors that very 
efficient dust destruction is unlikely in high-$z$ objects \citep{Dwek07, Gall11, Mattsson11}. It may of course be that dust destruction by SNe
is less efficient in high-$z$ galaxies, but also modelling of nearby late-type galaxies seems to work nicely without significant net destruction of 
dust \citep{Inoue03, Hirashita99}. In fact, no or little net dust destruction makes it easier to explain the dust-to-gas ratios, since stellar dust 
production is likely not sufficient for neither the Milky Way, nor any of the other late-type local group galaxies \citep{Hirashita99, Zhukovska08}.
 
Observational constraints imply dust production in SNe is rather inefficient \citep{Kotak06,Kotak09}, which suggest the high dust masses detected
in some, relatively old, SN remnants \citep[see, e.g.,][]{Morgan03,Morgan03b,Dunne09,Gall11b} could be the result of subsequent dust growth, {and/or 
heating of pre-existing dust,} rather than dust production in the actual SN. {However, since SN remnants with large dust masses
typically contain vast amounts of cold dust, some degree of growth appear to be necessary even if there is a component of heated swept-up dust.}
This {picture} is consistent with theoretical results which suggest 90\% of the dust produced in SNe is destroyed by the 
reverse shock before it reaches the ISM \citep{Bianchi07}. Hence AGB stars could be the most important source of stellar dust as a significant fraction
of the metals expelled by these stars is expected to be in the form of dust \citep{Edmunds98, Ferrarotti06}, which is supported by observational detections
of dust \citep[see, e.g., the recent results by][]{Ramstedt11}. It should be noted, however, that this 
picture may need to be revised due to the recent discovery of a large amount
of cold dust associated with SN 1987A \citep{Matsuura11}.

In models of dust evolution for the solar neighbourhood by \citet{Dwek98} and \citet{Zhukovska08}, the limited dust production in stars and possible 
dust destruction by SN-shock waves are more than well compensated by an efficient dust growth in the ISM, which is supported by observations 
indicating the existence of large, micrometer-sized dust grains in dense molecular clouds \citep{Pagani10}. There is further evidence from dust-evolution
modelling \citep[see, e.g.][]{Mattsson11,Pipino11,Valiante11} along with some observational constraints \citep[see, e.g.][]{Michalowski10} which 
suggests a need for significant dust growth in the early Universe as well. 
Dust growth 
appears to dominate over dust destruction also in the local, present-day Universe \citep{Hirashita99,Inoue03,Hirashita11,Asano11}. It is difficult to separate one 
scenario where dust growth is totally dominating over dust destruction from another where there is very little dust destruction and less dust growth.
But whether there is {\it net} growth/destruction it should have observable consequences, however.

We propose here a diagnostic tool for determining whether there is net dust growth or net dust destruction in the ISM of a late-type galaxy for which 
dust-to-gas as well as metallicity gradients can be derived. As we will show in this paper, the change (gradient) of the dust-to-metals ratio along a
galactic disc is closely connected to growth and destruction of dust in the ISM. In an associated paper \citep[][hereafter cited as Paper II]{Mattsson11c},
we investigate the implications of observed dust-to-metals profiles in a selection of late-type galaxies from the SINGS \citep{Kennicutt03} sample.

\section{Basic equations}
\label{theory}
In order to obtain analytical solutions and be able to manipulate the basic equations of the dust-enrichment problem in such a way that we can derive
some basic constraints, we use the instantaneous recycling approximation \cite[IRA, which essentially means all stars are assumed to have negligible 
lifetimes {with respect to the overall time scale for the build-up of metals and dust}, see][]{Pagel97} throughout this paper. No delayed element production 
due to stellar lifetimes is considered. 

For convenience, we also define the dust destruction rate relative to the growth rate of the stellar component $\Sigma_{\rm s}$ as
\begin{equation}
D(r,t) \equiv \dot{\Sigma}_{\rm ISM}(r,t)\left({d\Sigma_{\rm s}\over dt}\right)^{-1},
\end{equation}
where $\dot{\Sigma}_{\rm ISM}$ is the dust destruction rate due to SNe {and the variables $r$, $t$ are the galactocentric distance and time/age respectively
(a notation that we will assume is understood in the following)}. Similarly, we also define
\begin{equation}
G(r,t) \equiv \dot{\Sigma}_{\rm gr}(r,t)\left({d\Sigma_{\rm s}\over dt}\right)^{-1},
\end{equation}
where $\dot{\Sigma}_{\rm gr}$ is the rate of grain growth (in mass units) in the ISM.

Assuming a 'closed box', {where dust destruction in the ISM} is from SN-shocks, the equations for the metallicity $Z$ and the dust-to-gas ratio $Z_{\rm d}$ becomes
\begin{equation}
\label{metals}
\Sigma_{\rm g}{d Z \over d t} = y_{Z}{d\Sigma_{\rm s}\over dt} = -y_{Z}{d\Sigma_{\rm g}\over dt},
\end{equation}
\begin{equation}
\label{dust}
\Sigma_{\rm g}{d Z_{\rm d} \over d t} = y_{\rm d}{d\Sigma_{\rm s}\over dt} + Z_{\rm d}(r,t)\,[G(r,t)-D(r,t)]\,{d\Sigma_{\rm s}\over dt},
\end{equation}
{where $\Sigma_{\rm g}$ is the gas density, $\Sigma_{\rm d}$ is the dust density}, and the yield $y_{i}$ is defined as
\begin{equation}
\label{yield}
y_{i} = {1\over \alpha} \int_{m_{\rm lo}}^{m_{\rm up}} p_{i}(m)\,m\,\phi(m)\,dm,
\end{equation}
for both metals ($i = Z$) and stellar dust ($i = {\rm d}$).
In equation (\ref{yield}) above, $p_{i}$ is the fraction of the initial mass $m$ of a 
star ejected in the form of newly produced metals or dust, {$\alpha$ is the stellar lock-up fraction (i.e., the fraction of the baryon mass being locked up in long lived
stars)} and $\phi(m)$ is the mass-normalised IMF, {with $m_{\rm lo}$, $m_{\rm up}$ being the lower and upper mass cuts, respectively.}
Combining equations (\ref{metals}) and (\ref{dust}), we have
\begin{equation}
\label{dustzeqn}
{\partial Z_{\rm d}\over \partial Z} = {y_{\rm d}+Z_{\rm d}[G(r,t)-D(r,t)]\over y_Z},
\end{equation}
which thus have no explicit dependence on the gas mass density $\Sigma_{\rm g}$ or the stellar mass density $\Sigma_{\rm s}$, {although $G$, $D$ and $Z_{\rm d}$
of course may have implicit dependencies on the amount of gas and stars being present in a certain galactic environment}.

\section{Constraints on dust-to-metals gradients}
\label{constraints}
We will now prove some basic properties of dust-to-metals ($\zeta$) gradients relative to the metallicity ($Z$) gradient. 
For 'logarithmic' dust-to-metals and metallicity gradients we use the following notations,
\begin{equation}
\Delta_{Z_{\rm d}} \equiv {\partial\ln Z_{\rm d}\over \partial r} = {1\over Z_{\rm d}}{\partial Z_{\rm d}\over \partial r}, \quad 
\Delta_Z \equiv {\partial\ln Z\over \partial r} = {1\over Z}{\partial Z\over \partial r},
\end{equation}
\begin{equation}
\label{Deltazeta}
\Delta_{\zeta} \equiv {\partial\ln(Z_{\rm d}/Z)\over \partial r} =  {\partial\ln Z_{\rm d}\over \partial r} - {\partial\ln Z\over \partial r} =  {1\over Z_{\rm d}}{\partial Z_{\rm d}\over \partial r}-{1\over Z}{\partial Z\over \partial r},
\end{equation}
which are used since they both have the same unit ($[{\rm length}]^{-1}$). The two gradients $\Delta_\zeta$ and $\Delta_Z$ can be regarded as coupled through a function $f$ which may be seen as a function of a
number of physical parameters, but in general we may say it is a function of time $t$ and radial position (galactocentric distance) $r$ along the galaxy. Hence, we consider a relation of the form 
$\Delta_\zeta(r,t) = f(r,t)\,\Delta_Z(r,t)$. In the following we will implicitly assume all quantities except $y_{\rm d}$, $y_Z$ are functions of $r$ and $t$. We will also refer to the case of a zero derivative with respect to 
$r$ as a 'flat' gradient, which of course could be seen as the case of no gradient. However, we prefer to describe the gradients as being either positive, flat or negative,
{where 'negative' refers to a gradient (derivative) which decreases with galactocentric radius and vice versa for 'positve' gradients. Below we also use the sign function
${\rm sgn}(x)\equiv x/|x|$ to denote the sign of $\Delta_\zeta$ and $\Delta_Z$.}
\\[3mm]
{\bf THEOREM.} 
For a closed-box model, without any pre-enrichment, and where the IRA and constant yields $y_Z$, $y_{\rm d}$ have been adopted, the following always hold:
\begin{enumerate}
\item A flat (no slope) dust-to-metals gradient can only be obtained if there is neither net growth, nor any net destruction of dust in the ISM ($G=D$) or if the metallicity gradient is flat.
\item If the dust-to-metals and metallicity gradients have the same sign, there has to be net growth ($G>D$) of dust in the ISM.
\item If the dust-to-metals and metallicity gradients have opposite signs, there has to be net destruction ($G<D$) of dust in the ISM.
\end{enumerate}
\begin{flushright}
{$\square$}
\end{flushright}
{\it Proof.}  From the basic equations of dust evolution (see section \ref{theory}) one finds
\begin{equation}
\label{zetaeqn}
Z{\partial \zeta\over \partial Z} = {y_{\rm d}\over y_Z }+\left[(G-D) {Z\over y_Z} -1\right]\,\zeta.
\end{equation}
By use of the chain rule\footnote{If $\zeta$ is a function of $r$ and $t$ with continuous first partial derivatives, and if $r$ and $t$ can be regarded as differentiable functions of $Z$, then
\begin{displaymath}
{d\zeta\over dZ} = {\partial\zeta\over\partial r}{dr\over dZ}+{\partial\zeta\over\partial t}{dt\over dZ}.
\end{displaymath}
At a specific time $t = t_0$ we can thus write
\begin{displaymath}
{d\zeta\over dZ} = {d\zeta\over dr}{dr\over dZ}.
\end{displaymath}}, we get 
\begin{equation}
{d \zeta\over dr} = \left[{y_{\rm d}\over y_Z}{1\over Z} + \zeta\left({G-D\over y_Z} -{1\over Z}\right)\right] {d Z\over dr} ,
\end{equation}
which in terms of $\Delta_{\zeta}$ and $\Delta_Z$, can be written as
\begin{equation}
\label{sune}
\Delta_{\zeta} = \left[{y_{\rm d}\over y_Z}{1\over\zeta}+{Z\over y_Z}(G-D)-1\right]\Delta_Z.
\end{equation}
The function $f$ (see definition above) is then generally expressed
\begin{equation}
f \equiv {y_{\rm d}\over y_Z}{1\over\zeta}+{Z\over y_Z}(G-D)-1.
\end{equation}

\begin{enumerate}
\item
If $\Delta_Z=0$, then obviously $\Delta_\zeta=0$ as a consequence of Equation (\ref{sune}).
In case there is neither net growth, nor any net destruction of dust in the ISM ($G=D$), we have
\begin{equation}
\label{sune2}
f = {y_{\rm d}\over y_Z}{1\over\zeta}-1.
\end{equation}
Equation (\ref{dustzeqn}) gives
\begin{equation}
{\partial Z_{\rm d}\over \partial Z} = {y_{\rm d}\over y_Z } ,
\end{equation}
and again by the chain rule,
\begin{equation}
\label{flatgrad}
\left({d Z_{\rm d}\over dr}\right)_{G=D} = {y_{\rm d}\over y_Z}{d Z \over dr}.
\end{equation}
Integrating equation (\ref{flatgrad}), together with the natural initial conditions $Z(r,0)=Z_{\rm d}(r,0)=0$ (no pre-enrichment), one obtains $\zeta = {y_{\rm d}/ y_Z}$, or
\begin{equation}
{y_{\rm d}\over y_Z}{1\over\zeta}=1.
\end{equation}
Hence, according to Equation (\ref{sune2}), we must have $\Delta_\zeta = 0$, since $f = 0$, which proves part (i).\\

\item
First, we note that if ${\rm sgn}(\Delta_\zeta) = {\rm sgn}(\Delta_Z)$, then $f >0$. 
In case $G>D$, Equation (\ref{dustzeqn}) gives
\begin{equation}
\label{samesign}
\left({dZ_{\rm d}\over dr}\right)_{G>D} > \left({d Z_{\rm d}\over dr}\right)_{G=D}. 
\end{equation}
Then, by Equation (\ref{Deltazeta}) and the fact that $f=0$ if $G=D$, we have $\Delta_{\zeta,\,G>D} > \Delta_{\zeta,\,G=D}=0$, which implies $f>0$.
In case $G<D$, Equation (\ref{dustzeqn}) gives
\begin{equation}
\label{samesign}
\left({d Z_{\rm d}\over dr}\right)_{G<D} < \left({d Z_{\rm d}\over dr}\right)_{G=D}. 
\end{equation}
Again, using Equation (\ref{Deltazeta}) and $f=0$ if $G=D$, we have $\Delta_{\zeta,\,G<D} < \Delta_{\zeta,\,G=D}=0$, which implies $f<0$.
Hence, ${\rm sgn}(\Delta_\zeta) = {\rm sgn}(\Delta_Z)$ is possible if (and only if) $G>D$, which proves part (ii).\\

\item
In this case, if ${\rm sgn}(\Delta_\zeta) \ne {\rm sgn}(\Delta_Z)$, then $f<0$.
If $G>D$, Equation (\ref{dustzeqn}) gives
\begin{equation}
\label{samesign}
\left({dZ_{\rm d}\over dr}\right)_{G>D} > \left({d Z_{\rm d}\over dr}\right)_{G=D}. 
\end{equation}
Analogous to case (ii) we have $\Delta_{\zeta,\,G>D} > \Delta_{\zeta,\,G=D}=0$, which implies $f>0$.
In case $G<D$, Equation (\ref{dustzeqn}) gives
\begin{equation}
\label{oppsign}
\left({d Z_{\rm d}\over dr}\right)_{G<D} < \left({d Z_{\rm d}\over dr}\right)_{G=D}. 
\end{equation}
Thus, we have $\Delta_{\zeta,\,G<D} < \Delta_{\zeta,\,G=D}=0$, which implies $f<0$ and therfore
${\rm sgn}(\Delta_\zeta) \neq {\rm sgn}(\Delta_Z)$ is possible if (and only if)  $G<D$, which proves part (iii).
\end{enumerate}

\begin{flushright}
{$\square$}
\end{flushright}

\section{Simple models of dust growth and dust destruction}
\subsection{Dust growth in the ISM}
\label{growth}
The most likely dominant type of 'secondary' dust production is that by accretion of atoms (or small molecules) onto pre-existing interstellar dust grains. Dust grains can in
principle also grow by coagulation, but this process will not affect the total dust mass very much since it is mostly smaller dust grains being joined together into larger 
grains. Hence, we will here only discuss dust growth by accretion. 

We define the rate per unit volume at which the number of atoms $N_{\rm A}$ in dust grains 
grows by accretion of metals onto these dust grains in a similar way as \citep[see, e.g.][]{Dwek98}
\begin{equation}
{dN_{\rm A}\over dt} = f_{\rm s}\,\pi a^2 n_Z n_{\rm gr} \langle v_{\rm g}\rangle,
\end{equation}
where $n_Z$ and $n_{\rm gr}$ are the total {atomic} metals and dust-grain number densities {in the ISM}, respectively, $a$ is the typical grain radius and $f_{\rm s}$
is the sticking coefficient (i.e., the probability that an atom will stick to the grain). $\langle v_{\rm g}\rangle$ is the mean thermal speed of the gas particles (including 
metals), which is defined as
\begin{equation}
\langle v_{\rm g}\rangle \equiv \int_0^{\infty} v \, f(v) \, dv= \sqrt { \frac{8kT}{\pi m_{\rm A}}},
\end{equation}
where $f(v)$ is the Maxwell distribution, $k$ is the Boltzmann constant, $T$ is the kinetic temperature of the gas and $m_{\rm A}$ is the atomic weight of the gas 
particles. In terms of surface densities in the molecular gas clouds where the dust may grow, we can write
\begin{equation}
{d\Sigma_{\rm d}\over dt} = {f_{\rm s}\,\pi a^2 \tilde{\Sigma}_Z \Sigma_{\rm d} \langle v_{\rm g}\rangle \over \langle m_{\rm gr}\rangle d_{\rm c}},
\end{equation}
where {$\tilde{\Sigma}_Z$ is the surface density of free (atomic) metals}, $\langle m_{\rm gr}\rangle$ is the mean {mass of the dust grains} in the ISM {and 
$d_{\rm c}$ is the size of the molecular cloud in which the dust is growing}. The timescale of grain growth can then be expressed as
\begin{equation}
\label{taugr}
\tau_{\rm gr} = \tau_0\left(1-{Z_{\rm d}\over Z}\right)^{-1},
\end{equation}
where
\begin{equation}
\tau_0  = {\langle m_{\rm gr}\rangle\, d_{\rm c}\over f_{\rm s}\,\pi a^2\Sigma_Z  \langle v_{\rm g}\rangle}
              \approx {\langle m_{\rm gr}\rangle\, d_{\rm c}\over f_{\rm s}\,\pi a^2 Z\,\Sigma_{\rm mol} \langle v_{\rm g}\rangle},
\end{equation}
in which $\Sigma_{\rm mol} $ is the surface density of molecular gas, {and $Z$ the metallicity}. 

For simplicity we will assume $\Sigma_{\rm mol} \approx \Sigma_{\rm H_2} $, since most of the gas in the molecular gas
clouds is in the form of molecular hydrogen. We also assume $\Sigma_{\rm H_2} $ traces the star-formation rate, i.e.,
\begin{equation}
\dot{\Sigma}_\star = \eta\Sigma_{\rm H_2} = {1\over \alpha}{d\Sigma_{\rm s}\over dt},
\end{equation}
as indicated by several observational studies \citep[e.g.,][]{Rownd99,Wong02,Bigiel08,Leroy08,Bigiel11,Feldmann11,Schruba11}. Such a relation is also supported by theory and recent numerical 
experiments \citep[see, e.g.][]{Krumholz09,Krumholz11}. Moreover, the mean thermal speed $\langle v_{\rm g}\rangle$ is roughly constant in the considered ISM environment 
and the typical grain radius does not vary much. Hence, the timescale $\tau_0$ is essentially just a simple function of the metallicity, {the gas abundance} and the 
growth rate of the stellar component,
\begin{equation}
\tau_0 ^{-1}= {\epsilon Z \over\Sigma_{\rm g}} {d\Sigma_{\rm s}\over dt},
\end{equation}
the constant $\epsilon$ will, in the following, be treated as {an essentially free (but not unconstrained)} parameter of the model. The expected value is on the order of a 
few hundred, which is required to obtain $\tau_{\rm gr}\sim 10^{7}$ yr, suggested above. We will here adopt
\begin{equation}
\label{growthrate}
\left({dZ_{\rm d}\over dt}\right)_{\rm gr}  = \left(1-{Z_{\rm d}\over Z}\right) {Z_{\rm d}\over \tau_0} = {Z_{\rm d}\over \tau_{\rm gr}},
\end{equation}
as the rate of change of the dust-to-gas ratio $Z_{\rm d}$ due to accretion of metals onto pre-existing dust grains in the ISM. 
Note that this formulation of 'secondary' dust production differs from that used by \citet{Edmunds01} and \citet{Mattsson11} in that it also depends on the dust abundance in the ISM and the depletion of metals in atomic state.

\subsection{Dust destruction}
\label{dustdest}
The dominant mechanism for dust destruction is by sputtering in the high-velocity interstellar shocks driven by SNe, which can be directly related to the energy of the SNe \citep{Nozawa06}.
Following \citet{McKee89,Dwek07} the dust destruction time-scale is
\begin{equation}
\tau_{\rm d} = {\Sigma_{\rm g}\over \langle m_{\rm ISM}\rangle\,R_{\rm SN}},
\end{equation}
where $\Sigma_{\rm g}$ is the gas mass density, $\langle m_{\rm ISM}\rangle$ is the effective gas mass cleared of dust by each SN event, and $R_{\rm SN}$ is the SN rate, which may be approximated as
\begin{equation}
\label{snr}
R_{\rm SN}(t) \approx \dot{\Sigma}_{\rm s}(r,t)\int_{8M_\odot}^{100M_\odot} \phi(m)\,dm. 
\end{equation}
The integral in equation (\ref{snr}) is a constant with respect to time, and is not likely to vary
much over the disc either, hence the time scale $\tau_{\rm d}$ may be expressed as
\begin{equation}
\label{taud}
\tau_{\rm d}^{-1} \approx  {\delta\over \Sigma_{\rm g}}{d\Sigma_{\rm s}\over dt},
\end{equation}
where $\delta$ will be referred to as the dust destruction parameter. This parameter is dimensionless, and as such it can be seen as a measure of the efficiency of
dust destruction. More precisely, however, the efficiency is set by the fraction $f_{\rm d}$ of interstellar dust destroyed in an encounter with a SN shock wave, which 
occurs in the definition of $\langle m_{\rm ISM}\rangle$ \citep{McKee89,Dwek07},
\begin{equation}
\langle m_{\rm ISM}\rangle \equiv \int_{v_0}^{v_{\rm f}} f_{\rm d}(v_{\rm s})\left | {dM_{\rm sw}\over dv_{\rm s}}\right|\,dv_{\rm s},
\end{equation}
where $M_{\rm sw}$ is the swept-up gas mass (during Sedov-Taylor expansion), $v_{\rm s}$ is the shock velocity, and $v_0$, $v_{\rm f}$ are the initial and the final
velocity, respectively. Note that in this way $\delta$ is similar to the $\bar{\epsilon}$-parameter (average grain-destruction efficiency) used by \citep{McKee89}, which
should not be confused with the $\epsilon$ (dust-growth parameter) introduced in the previous section. {It should also be stressed that $f_{\rm d}$ is not a constant, but
a function of the shock velocity $v_{\rm s}$.}

A \citet{Larson98} IMF and $\langle m_{\rm ISM} \rangle \sim 1000 M_\odot$ \citep{Dwek07} suggests $\delta \sim 10$, {which is likely close to an upper limit for
$\delta$. Just as in the case of $\epsilon$ above, it is not absolutely clear, however, that $\delta$ can be treated as a parameter that does not vary during the course of
evolution of the ISM in a galaxy, but it seems in a given environment a fair approximation.

\section{Analytic solutions}
\label{models}
For simplicity we have assumed a closed box (see section \ref{constraints}), i.e., no in- or outflows to/from the disc. This is not in agreement with the widely accepted ideas about galaxy-disc 
formation, where the baryons (in
the form of essentially pristine gas) are assumed to be accreted over an extended period of time. But as shown by \citet{Edmunds90}, the only major effect of unenriched infall is to make the
effective yield smaller, i.e., to dilute the gas so that the metallicity builds up more slowly. As we in this study uses the present-day metallicity as input, the overall effects
of assuming a closed box are rather small, and in general only accretion of metal-enriched gas can affect the dust-to-metals ratio significantly (see Appendix \ref{infall}). 

\subsection{General solution}
Adopting the closed-box scenario, the dust destruction and dust growth models as described above, results in an equation for dust evolution,
\begin{equation}
\label{modeq}
\Sigma_{\rm g}{dZ_{\rm d}\over dt} = \left\{y_{\rm d} + Z_{\rm d}\left[\epsilon \left(1-{Z_{\rm d}\over Z} \right)\,Z -\delta\right]\right\}\,{d\Sigma_{\rm s}\over dt},
\end{equation}
which combined with the metallicity $Z$ gives
\begin{equation}
\label{dustz}
{dZ_{\rm d}\over dZ} = {1\over y_Z}\left\{y_{\rm d} + Z_{\rm d} \left[\epsilon \left(1-{Z_{\rm d}\over Z} \right)\,Z -\delta\right] \right\},
\end{equation}
where $y_Z$ is the metal yield. Provided $y_{\rm d}< y_Z$, the general closed-box solution (of equation \ref{dustz}) for the dust-to-gas ratio $Z_{\rm d}$ in terms of the metallicity $Z$ is (see
Appendix \ref{hypergeo} for a sketchy derivation),
\begin{equation}
\label{gensol}
Z_{\rm d} = {y_{\rm d}\over y_Z}\left(Z-{\delta\over \epsilon}\right)
\left\{{
\varphi_{11}\left[{y_{\rm d}\over y_Z}, {(\epsilon Z - \delta)^2\over \epsilon y_Z}, {\delta^2\over \epsilon y_Z}\right] - 
\varphi_{11}\left[{y_{\rm d}\over y_Z}, {\delta^2\over \epsilon y_Z}, {(\epsilon Z - \delta)^2\over \epsilon y_Z}\right] 
\over 
2\varphi_{10}\left[{y_{\rm d}\over y_Z}, {\delta^2\over \epsilon y_Z}, {(\epsilon Z - \delta)^2\over \epsilon y_Z}\right] + 
\varphi_{01}\left[{y_{\rm d}\over y_Z}, {(\epsilon Z - \delta)^2\over \epsilon y_Z}, {\delta^2\over \epsilon y_Z}\right] 
}\right\} 
\end{equation}
for 
\begin{equation}
\varphi_{ij}(k,x,y) \equiv M\left[i+{k\over2}, i+ {1\over2};{x\over 2}\right]\,U\left[j+{k\over2}, j+{1\over2};{y\over 2}\right].
\end{equation}
The functions $M$ and $U$ are the confluent hypergeometric Kummer-Tricomi functions of the first and second kind, respectively \citep[][see also Appendix \ref{numerical}]{Kummer1837,Tricomi47}.

In the equations above, $y_{\rm d}$ and $y_Z$ are the stellar dust and metal yields, respectively, $\delta$ is the 'dust destruction parameter' (see section \ref{dustdest}) 
and $\epsilon$ is the 'grain-growth parameter' (see section \ref{growth}). equation (\ref{gensol}) is singular at $Z = \delta/\epsilon$, which means this general solution must be used with care.
It is relatively straight forward to implement the Kummer-Tricomi functions numerically (see Appendix \ref{numerical}), but there is a regular singularity at the origin in Kummer's equation
(to which $M$ and $U$ are linearly independent solutions) which can cause potential problems in the vicinity of $Z = \delta/\epsilon$.

\subsection{Special cases}
\label{specialcases}
The general solution presented above is obviously not simple to use in practice, not the least because of the singularity at $Z = \delta/\epsilon$.
However, in the special case $y_{\rm d}\to 0$ (negligable net contribution of dust from stars) the singularity can be removed and the solution expressed as
(see Appendix \ref{hypergeo})
\begin{equation}
\label{nostars}
{Z_{\rm d} \over Z_{\rm d,\,0}} = \exp\left(\xi^2\right) \left\{\exp\left(\xi_0^2\right)  + \eta_0 Z_{\rm d,\,0}\left[{\rm erfi}(\xi) - {\rm erfi}(\xi_0)\right] \right\}^{-1}
\end{equation}
with $Z_{\rm d,\,0}$ being the initial dust-to-metals ratio,
\begin{equation}
\xi \equiv \sqrt{{1\over 2}{(\epsilon Z-\delta)^2\over \epsilon y_Z}}, \quad \xi_0 \equiv \sqrt{{1\over 2}{(\epsilon Z_0-\delta)^2\over \epsilon y_Z}}, \quad \eta_0 \equiv \sqrt{{\pi\epsilon\over 2y_Z}},
\end{equation}
and $Z_0$ the initial metallicity. In the solution above, ${\rm erfi}(z)$ is the imaginary error function, related to the ordinary error function ${\rm erf}(z)$ as ${\rm erfi}(z) = -i\,{\rm erf}(i\,z)$, where 
${\rm erf}(z)$ is defined as
\begin{equation}
{\rm erf}(x) = \frac{2}{\sqrt{\pi}}\int_{0}^x e^{-t^2} dt.
\end{equation}
Physically, one may interpret this solution as describing the subsequent evolution (where the dust contribution from stars may be considered negligible) after an initial phase of metal enrichment 
and dust formation leading up to the point where $Z=Z_0$ and $Z_{\rm d = }Z_{\rm d,\,0}$. It may not be entirely realistic, but it demonstrates the interstellar "battle" between growth and destruction 
of dust grains in a very nice way (see Section \ref{growthvsdest} and Figure \ref{epsilondelta}).

Even when there is a significant net contribution from stars, we can still find simpler solutions for special cases.
In case there is no dust destruction by SNe ($\delta = 0$) the solution reduces to
\begin{equation}
Z_{\rm d} = {y_{\rm d}\over y_Z} {
M\left(1+{1\over 2}{y_{\rm d}\over y_Z}, {3\over 2}; {1\over 2}{\epsilon Z^2\over y_Z}\right)
\over 
M\left({1\over 2}{y_{\rm d}\over y_Z}, {1\over 2}; {1\over 2}{\epsilon Z^2 \over y_Z}\right)}
\,Z,
\end{equation}
and if there is dust destruction, but no grain growth in the ISM ($\epsilon = 0$), then
\begin{equation}
\label{nogrow}
Z_{\rm d} = {y_{\rm d} \over \delta}\left[1-\exp\left(-\delta{Z\over y_Z}\right)\right].
\end{equation}
If there is neither growth, nor destruction of dust in the ISM ($\epsilon = \delta = 0$), we have the trivial case
\begin{equation}
Z_{\rm d} = {y_{\rm d} \over y_Z} Z,
\end{equation}
corresponding to pure stellar dust production and obviously a flat dust-to-metals gradient. All the special cases above evade the inconvenient singularity at $Z = \delta/\epsilon$.

\section{Graphic analysis}
Using the numerical implementation of $M$ and $U$ described in Appendix \ref{numerical} we will here demonstrate the general behaviour of the dust-to-metals ratio $\zeta = Z_{\rm d}/Z$ using contour plots. 
Unless anything else is stated, we assume $y_Z = 0.02$ is a good typical metal yield \citep[which is consistent with the results of Paper II, but note that, e.g.,][finds a lower value]{Garnett02} and  
$y_{\rm d}={1\over 2}y_Z$ for simplicity. 

\subsection{General effects of growth destruction of dust in the ISM}
In case of no dust destruction ($\delta = 0$) the dust-to-metals ratio builds up to a maximum (where $\zeta\sim 1$) as $\epsilon$ and the metallicity
$Z$ increases (see Figure \ref{epsilon}, left panel). At low metallicities (half of solar or less, in the present case) the effect of increasing $\epsilon$ is relatively small once we get beyond a certain $\epsilon$, while at 
higher metallicities $\zeta$ grows rapidly until the metals reservoir is exhausted and $\zeta$ approaches unity \citep[as also found in the models by, e.g.,][]{Hirashita11,Asano11}. In case of no dust growth ($\epsilon = 0$) the dust-to-metals ratio is on a steep "downhill slope" 
(approaching $\zeta = 0$) for essentially all metallicities and $\delta$-values on the considered interval (see figure \ref{delta}, left panel). Note that $\zeta$ is very small at high metallicity if there is significant dust
destruction. 

\subsection{Dust-to-metals gradients}
As shown by the theorem in section \ref{constraints} the effect of dust destruction and dust growth on the dust-to-metals gradient in a galaxy disc is to make it steeper or flatter. The effect of growth and destruction of 
dust in the ISM can be illustrated in a more intuitive fashion if we consider the specific effects on a given metallicity profile. We here assume that metals in a disc follows an exponential distribution,
\begin{equation}
\label{expmetals}
Z(R) = Z_0 \exp\left(-{R\over R_0}\right),
\end{equation}
where we set the central metallicity to $Z_0 = 0.055$ and the $e$-folding scale length $R_0$ is set to be the unit for the galactocentric distance. The right panel of figure \ref{epsilon} shows how
dust growth creates a dust-to-metals gradient that falls of with galactocentric distance and becomes increasingly steeper as $\epsilon$ increases (for $\epsilon = 0$ the gradient is flat). Similarly, the right panel
of figure \ref{delta} shows how dust destruction creates an inwards gradient, starting from a flat gradient for $\delta = 0$. 

In the context of dust-to-metals gradients as signs of either net dust growth or net dust destruction, one should as well note that if the metallicity gradient and the dust-to-gas gradients are essentially flat, it is more 
or less impossible to distinguish between pure stellar dust production (albeit with a high stellar dust yield) and scenario including dust growth
and/or dust destruction in the ISM.

\subsection{Dust growths vs. destruction}
\label{growthvsdest}
Growth and destruction of dust must likely occur together. As we describe in Appendix \ref{hypergeo} the general solution with both $\epsilon$ and $\delta$ non-zero, has a singularity at $Z = \delta/\epsilon$,
which makes the analysis of how growth and destruction compete somewhat complicated and not least limited. Hence, we will here consider the special (and not entirely realistic) case where stellar dust
production is considered negligible (see equation \ref{nostars}) starting from a point in time when the metallicity $Z = 1.0\cdot 10^{-5}$ and the dust-to-metals ratio $\zeta = 0.5$. In figure \ref{epsilondelta} we
show $\zeta$ as a function of $\epsilon$ and $\delta$ for a fixed present-day metallicity $Z = 0.02$. Increasing the efficiency of dust destruction counteracts the dust growth,
which is shown by the "downhill slope" towards high $\delta$ and low $\epsilon$ values. Clearly, a high efficiency of dust destruction is not likely if there is to be a significant net production of dust without 
invoking a ridiculously short time scale for the dust growth in the ISM. More precisely, it is required that $\epsilon Z \gg \delta$, which in case $Z = 0.02$ and $\delta = 10$, would imply $\epsilon \gg 500$. With such a
large $\epsilon$ the typical growth-time scale is down to $\sim 10^5$ yr or less. As we mentioned in section \ref{growth}, $\epsilon$ should not exceed values of a few hundred if the dust-growth time scale $\tau_{\rm gr}$ 
is to be consistent with the suggested numbers for the local ISM of the Milky Way. It is quite possible that $\tau_{\rm gr}$ can be significantly shorter in, .e.g., a denser environment, but a
deeper analysis of this goes beyond the scope of this paper.

\subsection{"Critical" metallicity for dust growth}
Just as {\citet{Zhukovska08}, \citet{Hirashita11} and \citet{Asano11}}, we find that there exist a "critical metalliciy" $Z_{\rm crit}$ where the dust-mass contribution from grain growth increases rapidly. But this rapid increase over orders of magnitude
occurs only if the stellar dust yield $y_{\rm d}$ is significantly lower than the metal yield (see figure \ref{critz}, left upper panel).  Moreover, $Z_{\rm crit}$ depends somewhat on the dust-growth time scale (or $\epsilon$), 
which can be seen in figure \ref{critz} (right upper panel). Hence, $Z_{\rm crit}$ should not be viewed as a universal constant. In fact, a reasonable definition of $Z_{\rm crit}$ would be the metallicity at which stellar dust
production and the net dust growth in the ISM contribute equally to build-up of the interstellar dust component. In such a case, adopting the model used above, 
\begin{equation}
Z_{\rm crit} = Z_{\rm d} + {y_{\rm d}\over \epsilon Z_{\rm d}} - {\delta\over\epsilon}.
\end{equation}
If dust growth dominates over stellar dust production and dust destruction in the ISM is negligible, i.e., if $y_{\rm d}/\epsilon \ll 1$ and $\delta/\epsilon \ll 1$, then $Z_{\rm crit}  \approx Z_{\rm d}$, which suggest
 $Z_{\rm crit}  \sim \,y_{\rm d}/y_Z$. At metallicities below this value, the dust evolution (as function of metallicity) should be essentially identical to the case of pure stellar dust production - without any growth or
 destruction of dust grains in the ISM. In the right upper panel of figure \ref{critz}, $Z_{\rm crit} = y_{\rm d}/y_Z$ is marked by a vertical dashed line. At lower metallicities all model curves are indeed the same.

This critical metallicity $Z_{\rm crit}$ has an interesting implication for dust-to-metals/gas gradients, as it predicts the existence of bends also in logarithmic slopes 
and the existence of a critical galactocentric distance in between an inner and an outer "plateau" where $\zeta$ is constant (see Fig \ref{critz}, lower panels). This
non-linear feature is the consequence of the interstellar dust-growth rate (see equation \ref{growthrate}) having a non-linear ($Z_{\rm d}^2$) term. Although equation (\ref{taugr}) and equation (\ref{growthrate}) together represent 
one specific model, all models of interstellar dust growth will be non-linear as long as they depend on the amount of dust and "free" metals (not locked-up in dust) available.

    \begin{figure*}
  \resizebox{\hsize}{!}{
   \includegraphics{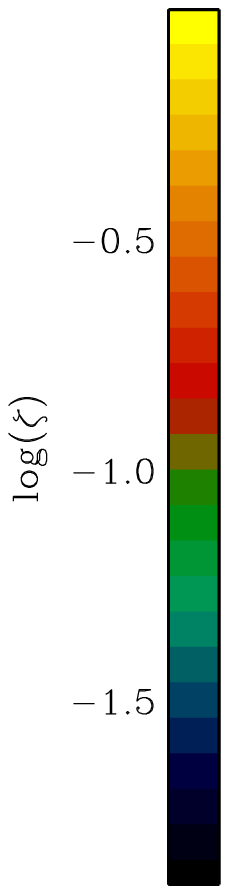}
   \includegraphics{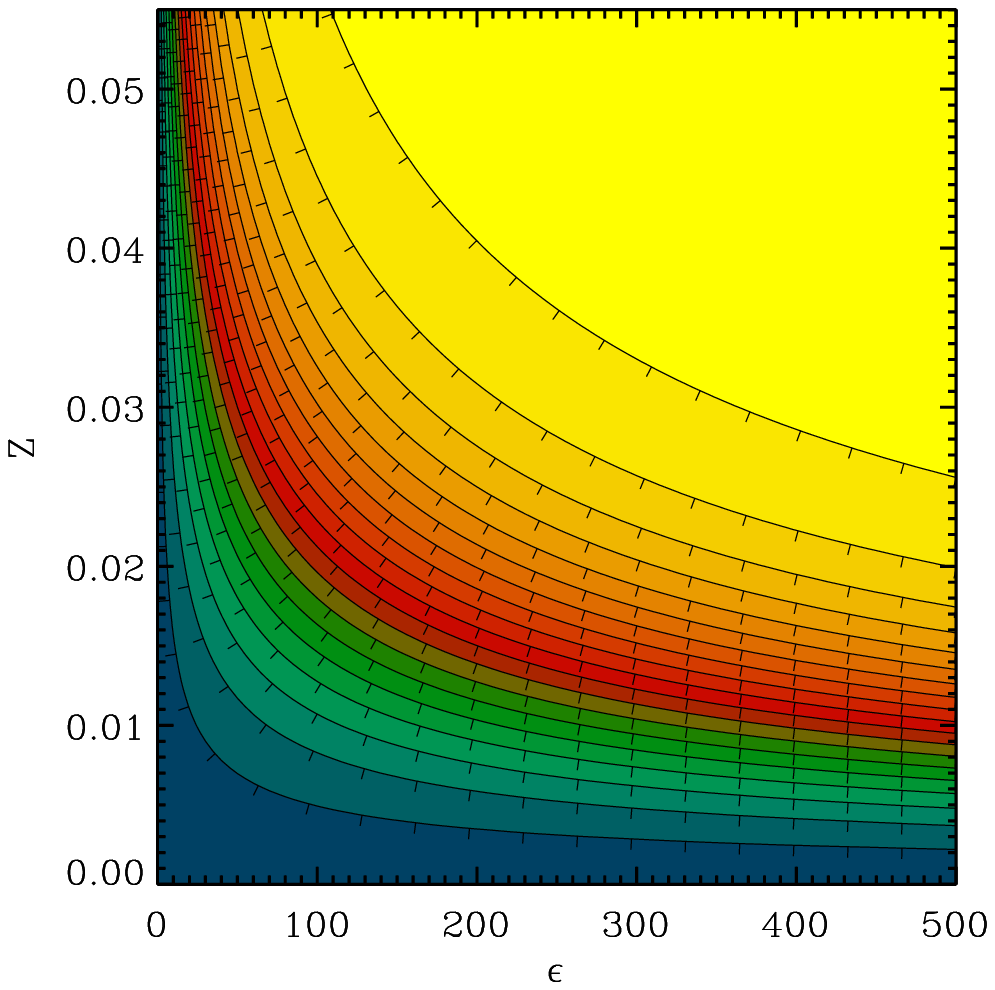}
   \includegraphics{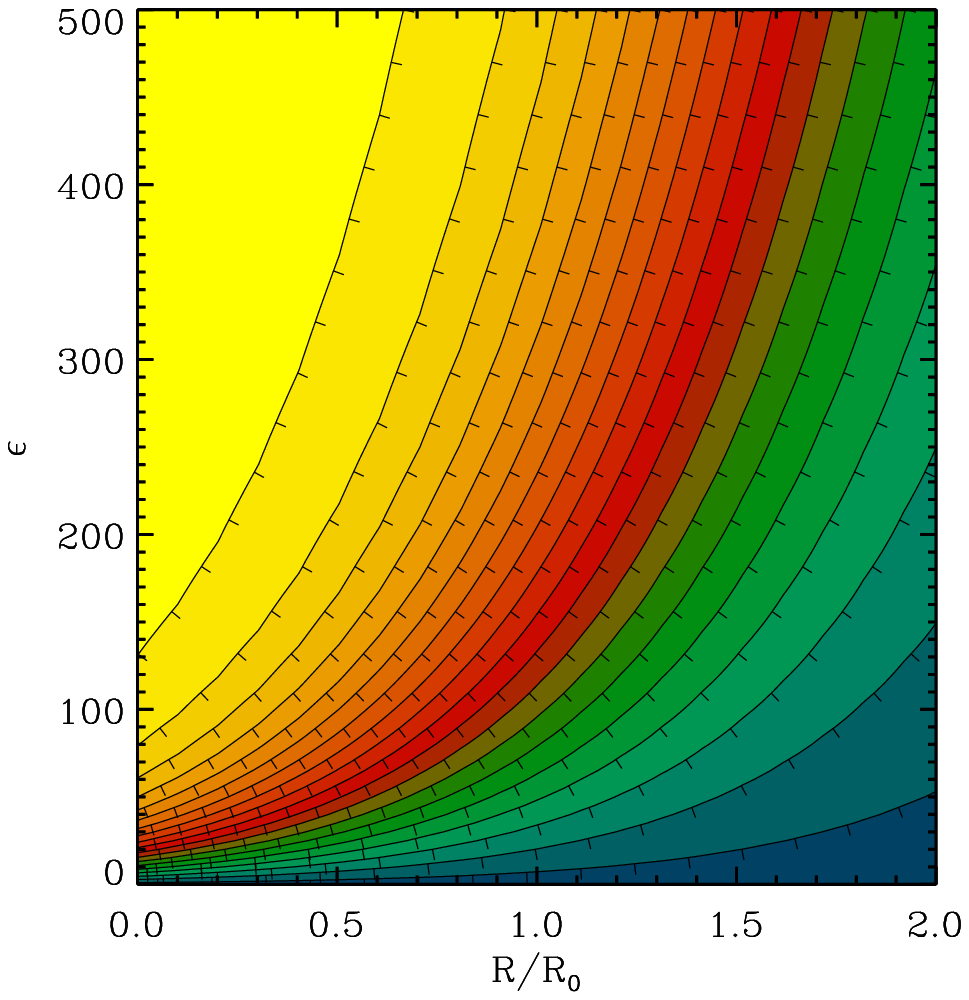}
  }
  \caption{\label{epsilon} Left: Dust-to-metals ratio $\zeta = Z_{\rm d}/Z$ as a function of the metallicity $Z$ and the dust-growth parameter $\epsilon$ for the case where there is no dust destruction due to SNe 
  ($\delta=0$). Right: Same as the left panel, but as a function of the galactocentric distance in a galaxy disc assuming an exponential distribution of metals.}
  \end{figure*}
  
      \begin{figure*}
  \resizebox{\hsize}{!}{
   \includegraphics{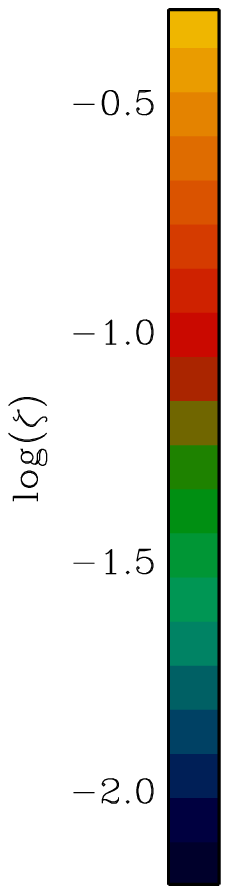}
   \includegraphics{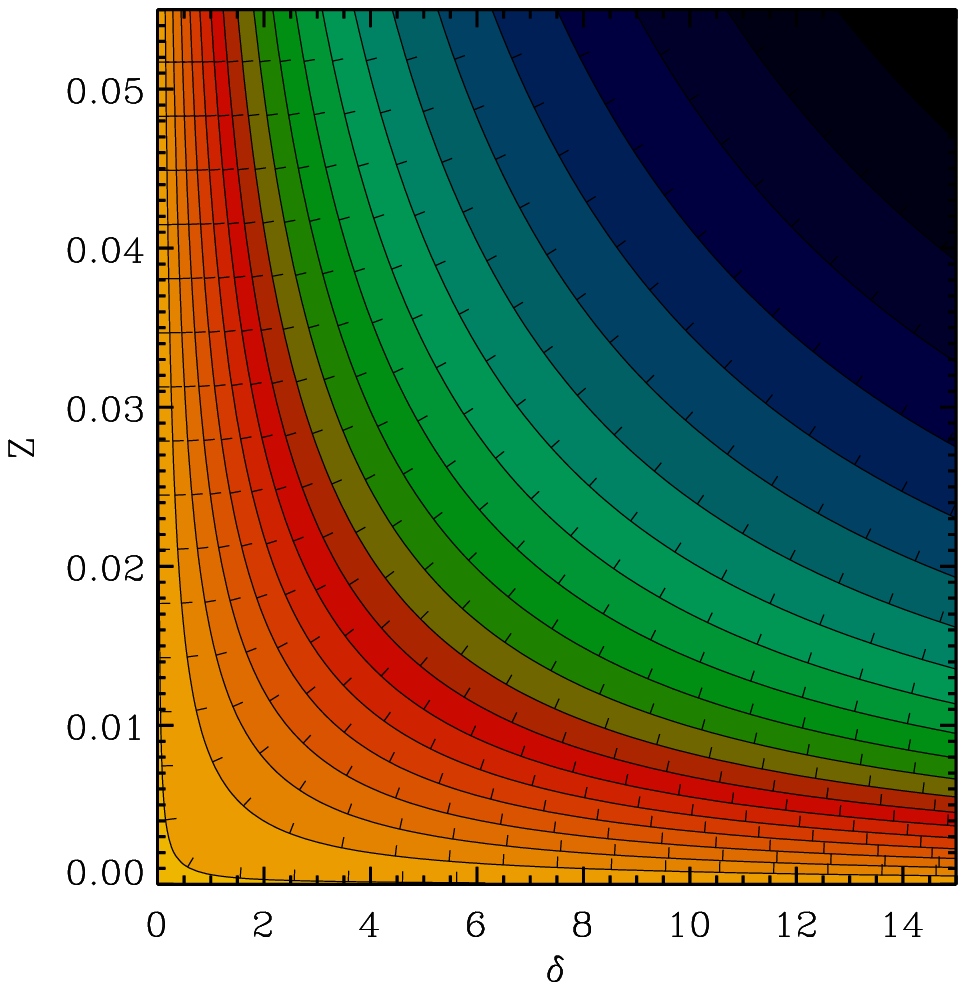}
   \includegraphics{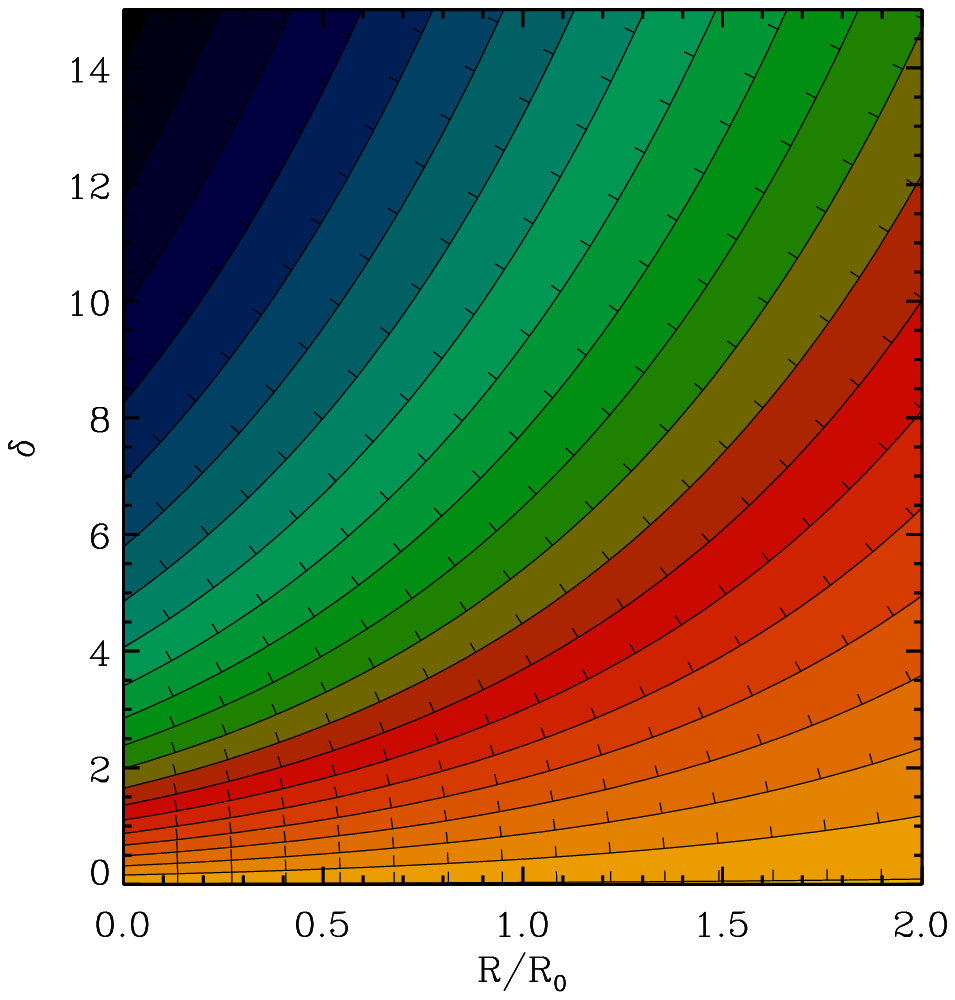}
  }
  \caption{\label{delta} Left: Dust-to-metals ratio $\zeta = Z_{\rm d}/Z$ as a function of the metallicity $Z$ and the dust-destruction parameter $\delta$ for the case where there is no dust growth in the ISM 
  ($\epsilon=0$). Right: Same as the left panel, but as a function of the galactocentric distance in a galaxy disc assuming an exponential distribution of metals.}
  \end{figure*}

      \begin{figure*}
  \resizebox{\hsize}{!}{
   \includegraphics{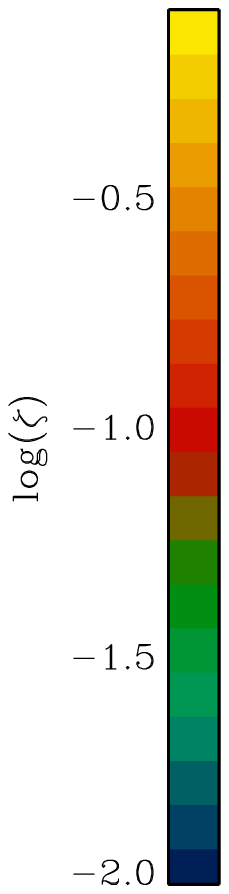}
   \includegraphics{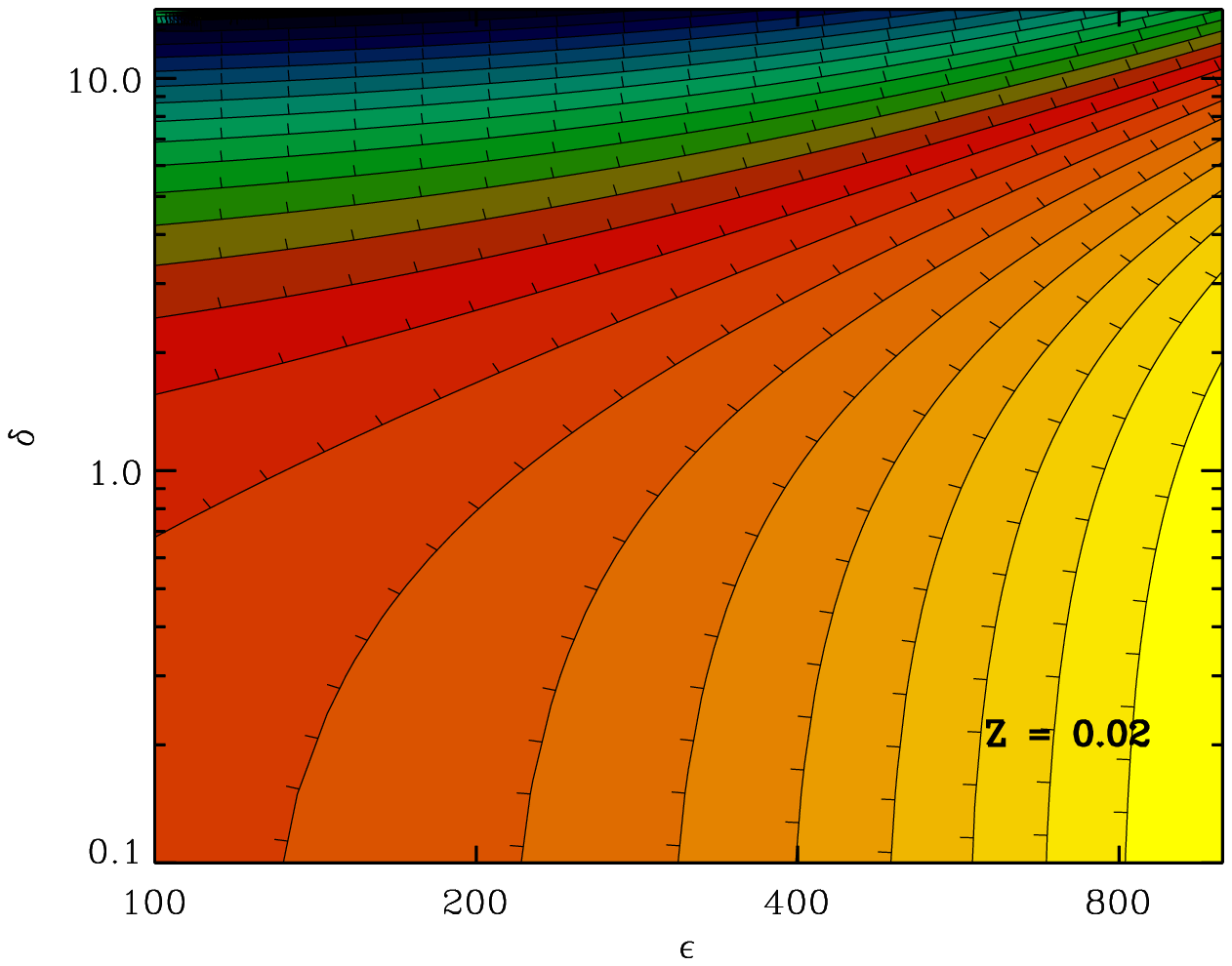}}
  \caption{\label{epsilondelta} 
  Contour plot of the dust-to-metals ratio as function of the parameters $\epsilon$ (growth) and $\delta$ (destruction) for the special (and not entirely realistic) case where stellar dust
  production is considered negligible (see equation \ref{nostars}) starting from a point in time when the metallicity $Z = 1.0\cdot 10^{-5}$ and the dust-to-metals ratio $\zeta = 0.5$. The present-day metallicity is assumed to be roughly solar ($Z = 0.02$).}
  \end{figure*}

        \begin{figure*}
  \resizebox{\hsize}{!}{
   \includegraphics{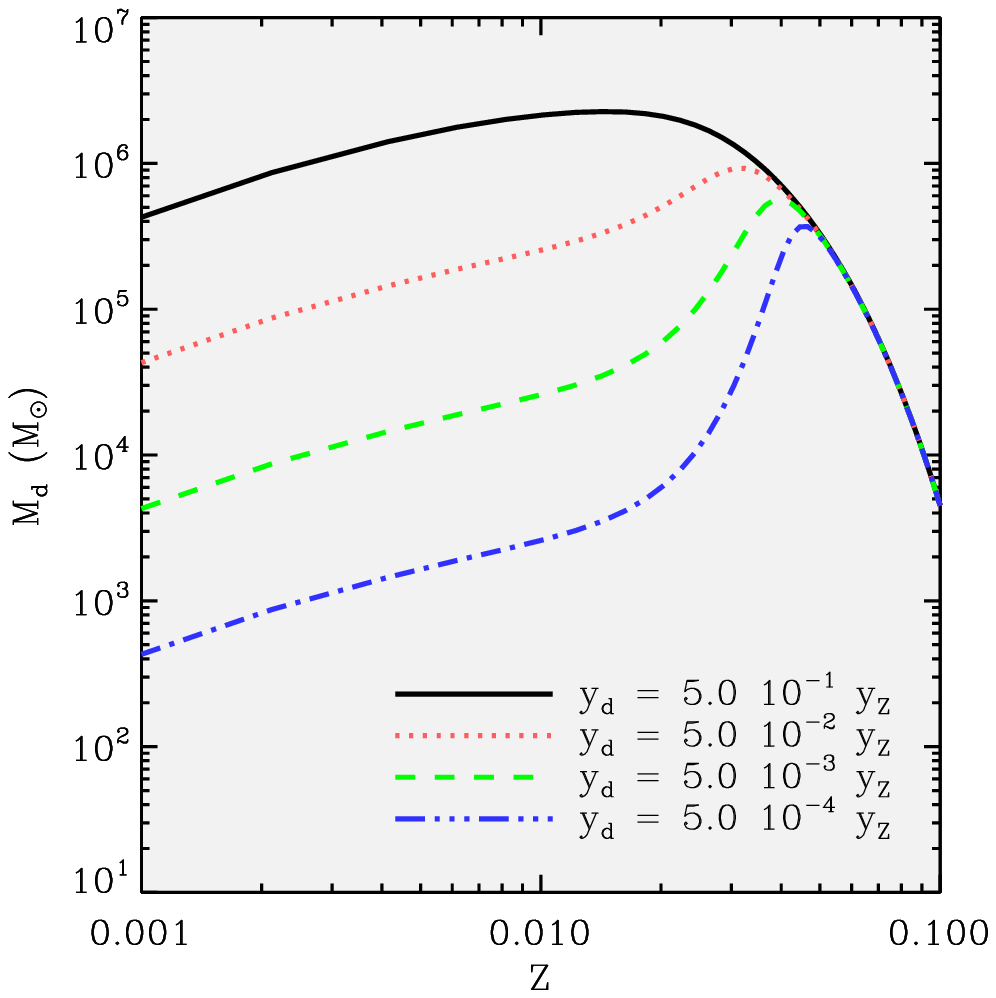}
   \includegraphics{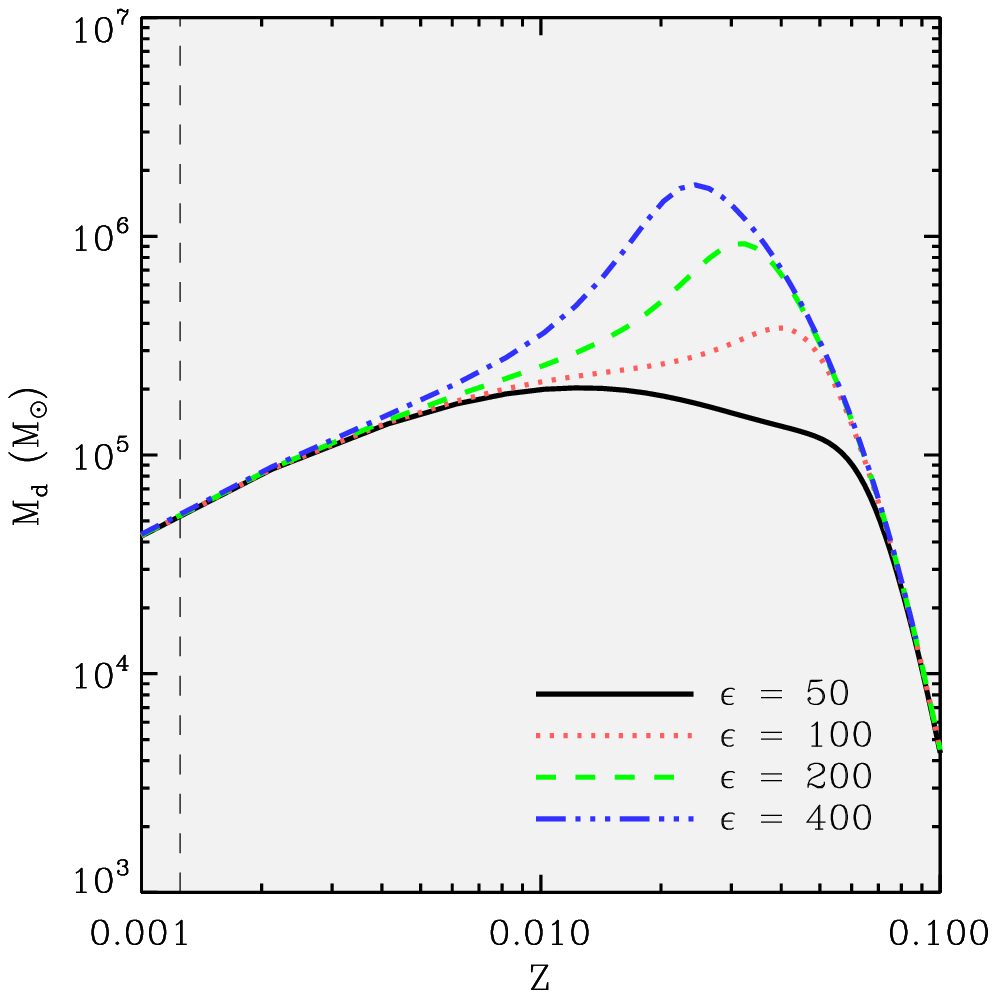}}
   \resizebox{\hsize}{!}{
   \includegraphics{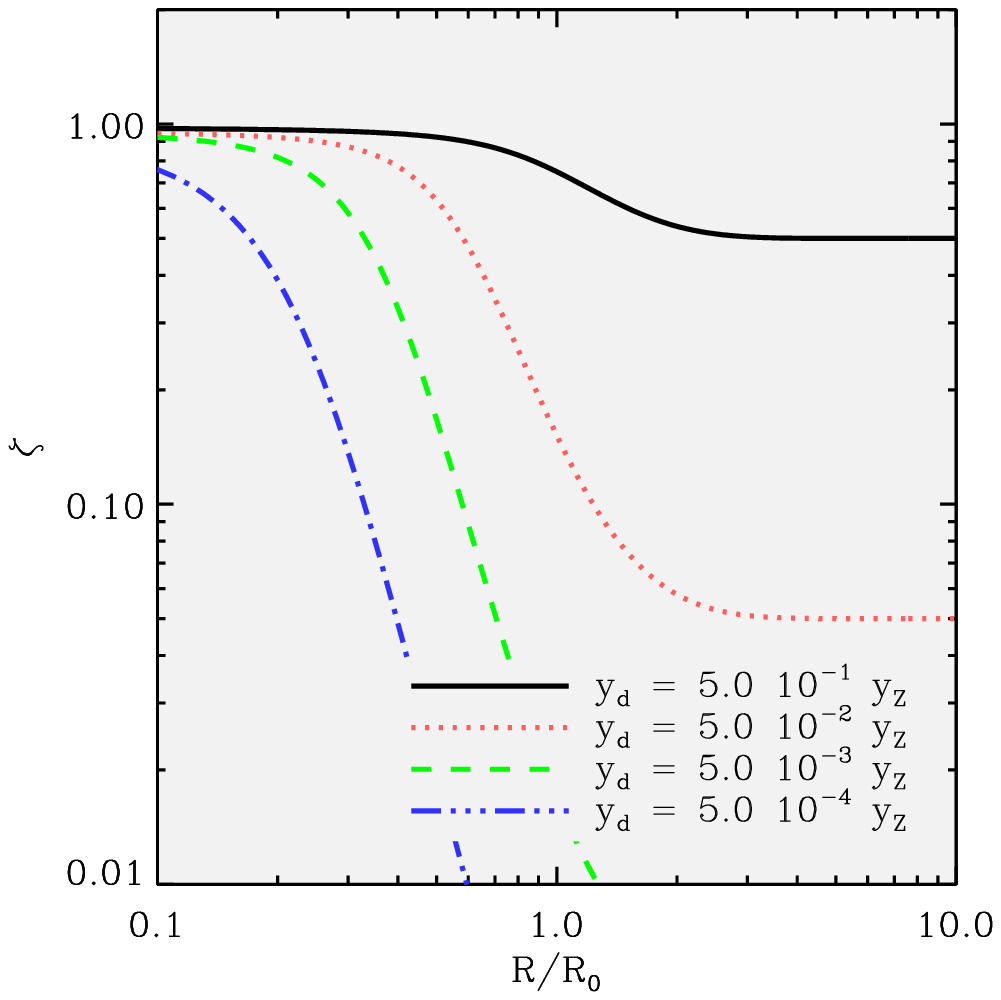}
   \includegraphics{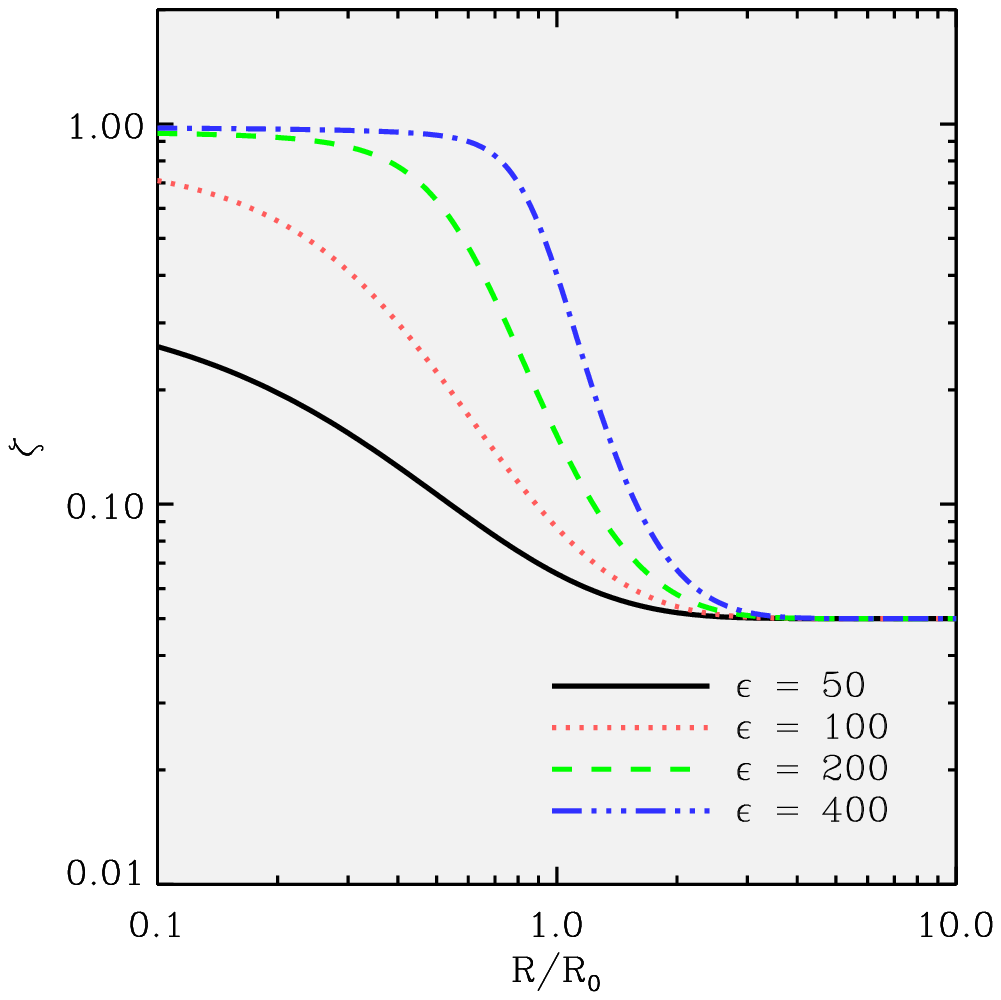}}
  \caption{\label{critz}
  Effects of the critical metallicity for dust growth domination and its dependence on the stellar dust yield and dust growth parameter $\epsilon$. The upper panels show the evolution of the dust mass as a function
  of metallicity for various values of the stellar dust yield $y_{\rm d}$ with a fixed $\epsilon = 200$ (left panel) and various values of $\epsilon$ with a fixed stellar dust yield $y_{\rm d} = 5.0\cdot 10^{-3} y_Z$ (right
  panel). The lower panels show the corresponding plots of the dust-to-metals ratio $\zeta$ as a function of galactocentric distance assuming an $e$-folding decay of the metallicity along the disc (see equation 
  \ref{expmetals}).}
  \end{figure*}

\section{Discussion and Conclusions}
We have shown that dust destruction by shock waves from exploding SNe and interstellar dust growth acts in opposite ways on the dust-to-metals gradient over a galaxy disc (see 
the theorem proved in section \ref{constraints}). This is hardly surprising, but starting from an exactly flat gradient (or no gradient, more precisely) dust destruction will over time 
create an inwards slope, while dust growth will create an outwards slope, provided the dust-to-gas ratio as well as the metallicity have negative gradients, i.e. decreases with 
galactocentric distance. Hence, we expect dust-to-metals gradients to have {\it positive} (inwards) gradients if dust destruction is more important than dust growth, and if dust growth 
is the more important process we expect them to be {\it negative} in general. The dust-to-metals gradient thus appears to be a useful diagnostic for the existence of interstellar dust 
growth. 

Our simple model of dust growth has just one adjustable parameter. This parameter ($\epsilon$) can have a rather wide range of numerical values depending on
what one assumes about the physical properties of the dust grains as well as the gas in ISM. In principle $\epsilon$ is proportional to the gas mass density 
$\Sigma_{\rm g}$ if all other quantities remain constant, but the star formation efficiency $\eta$ is likely proportional to $\Sigma_{\rm g}$ raised to some power 
 \citep{Krumholz05} and since the cloud size $d_{\rm c}$ is also likely related to $\Sigma_{\rm g}$, $\epsilon$ may not be much
dependent on $\Sigma_{\rm g}$ after all. More precisely, the star-formation efficiency (or time scale) is expected to correlate with the free-fall time scale, i.e., 
$\dot{\Sigma}_\star \propto \Sigma_{\rm H_2}/\tau_{\rm ff}$, where $\tau_{\rm ff} \propto \Sigma_{\rm g}^{-1/2}$ \citep{Krumholz09}, assuming
$\Sigma_{\rm g}\propto \rho_{\rm g}$ \citep{Elmegreen02}, and the scale of the cloud size $d_{\rm c}$ is given by Jeans length $\lambda_{\rm J}\propto \Sigma_{\rm g}^{-1/2}$. 
As $\epsilon \propto \tau_{\rm ff}\,d_{\rm c}\,\Sigma_{\rm g}$ and $\langle v_{\rm g}\rangle$ is roughly constant (isothermal conditions), the effective dependence on 
$\Sigma_{\rm g}$ is expected to be weak, if not negligible. Thus, it is fair to assume that $\epsilon$ is (effectively) only very weakly dependent on $\Sigma_{\rm g}$ {\it within} 
a galaxy, although from one galaxy to another $\epsilon$ may vary significantly, however (see Paper II). Below we analyse the range of possible 
$\epsilon$ values considering just mean/characteristic values of  $\Sigma_{\rm g}$, $\eta$ and $\langle v_{\rm g}\rangle$.
 
In terms of the included physical parameters (see section \ref{growth}), we find
\begin{equation}
\label{epsilondef}
\epsilon \approx {f_{\rm s} \pi a^2 \langle v_{\rm g}\rangle \over  \alpha \langle\eta\rangle \, \langle d_{\rm c}\rangle\, \langle m_{\rm gr}\rangle} \,\langle\Sigma_{\rm g}\rangle.
\end{equation}
The lock-up fraction $\alpha$ is 0.6 - 0.8 for a normal IMF \citep[we use here $\alpha = 0.7$, see][figure 1]{Mattsson11}, $\langle\eta\rangle$ is $\sim 1$ Gyr$^{-1}$ and since 
the typical size of a molecular cloud 
$d_{\rm c}$ is 10 - 100 pc, we adopt $\langle d_{\rm c} \rangle =50$~pc. The average grain mass $\langle m_{\rm gr}\rangle$ of course depends on the typical grain size 
$a$, where the latter ranges between $0.001 \mu$m for the smallest seed particles and $\sim 1 \mu$m for large full-grown dust grains. Hence, it is more convenient to 
introduce the characteristic grain density $\rho_{\rm gr} = \langle m_{\rm gr}\rangle/\langle V_{\rm gr}\rangle$, where $V_{\rm gr}$ is the volume of a dust grain. The grain 
density $\rho_{\rm gr}$ is typically $3.3$ g~cm$^{-3}$ \citep{Draine07} for silicates and $1.85$ g~cm$^{-3}$ for amorphous carbon \citep{Rouleau91}, but other values 
can also be found in the literature. Taking $\rho_{\rm gr} = 2.5$ g~cm$^{-3}$ as representative figure for cosmic dust in general, we arrive at
\begin{equation}
\epsilon \approx   4.2 \times\, f_{\rm s} \,\left({a\over \mu{\rm m}}\right)^{-1} \,\left({\langle\Sigma_{\rm g}\rangle \over M_\odot\,{\rm pc}^{-2}}\right).
\end{equation}
Assuming that all metals that come in contact with a dust grain will stick to that dust grain ($f_{\rm s} = 1$), a small characteristic grain size $a = 0.01 \mu$m and a relatively high
average gas density of $\Sigma_{\rm g} = 50\,M_\odot$~pc$^{-2}$,  will result in an $\epsilon$ of roughly $2\cdot 10^{4}$ 
corresponding to a typical grain-growth time scale of $\tau_{\rm gr} \sim 10^{6}$ yr if the gas consumption rate is similar to that of 
the solar neighbourhood. {Such high values of $\epsilon$ may be expected in young star-forming systems (e.g., late-type dwarf galaxies) where one has reasons to believe 
that gas densities are quite high and the grain-size distribution is biased towards small grains. The latter is due to insufficient time for extensive grain growth, and grain shattering 
owing to an elevated SN rate and strong UV radiation as consequences of recent star formation.} If $f_{\rm s} = 0.1$ (which is more consistent with silicate growth),  
$a = 1 \mu$m and $\Sigma_{\rm g} = 5\,M_\odot$~pc$^{-2}$, then 
$\epsilon \approx 2$ and thus $\tau_{\rm gr} \sim 10^9$ yr. With $f_{\rm s} = 1$, $a = 0.1 \mu$m and $\Sigma_{\rm g} = 10\,M_\odot$~pc$^{-2}$, the corresponding $\tau_{\rm gr}$ is 
$\sim 10^{8}$ yr assuming a gas-consumption rate and a metal content similar to that of the local Galaxy. This number is consistent with, estimates made in some other studies 
\citep{Jones96,Jones04,Zhukovska08,Jones11}, but slightly longer than the time scales suggested recently by \citet{Hirashita11}. Note that grain size is the parameter that is likely
most important for the value of $\epsilon$ as it can vary quite significantly. The gas mass density can vary over several orders of magnitude as well, but as
described above it may be cancelled out by other parameters.

The model of dust destruction due to SN shock waves has effectively only one parameter as well. This dust destruction parameter $\delta$ can be expressed as
\begin{equation}
\delta = {\langle m_{\rm ISM}\rangle\over \alpha} \,\int_{8M_\odot}^{100M_\odot} \phi(m)\,dm,
\end{equation}
where $\phi$ is the IMF and $\alpha$ is the lock-up fraction, as previously defined. With $\alpha = 0.7$ and a normal IMF \citep[see, e.g.,][]{Larson98}, we find
\begin{equation}
\delta \approx 0.018 \times \left(\langle m_{\rm ISM}\rangle \over M_\odot \right),
\end{equation}
which suggest $\delta$ is of order ten, if $\langle m_{\rm ISM}\rangle \sim 1000\,M_\odot$. The actual efficiency of dust destruction, and thus the effective interstellar gas 
mass cleared of dust, is not very well known. Therefore, it is reasonable to treat $\delta$ as an essentially free parameter. In order to have net growth of dust in the ISM, the 
value of $\delta$ needs to be $\delta < \epsilon\,(Z-Z_{\rm d})$. This means $\delta \sim 10$ is likely at the upper end of possible values for such a scenario, assuming
$Z-Z_{\rm d} \approx 0.01$, which suggest $\delta = 10$ would require $\epsilon > 1000$ or $\tau_{\rm gr} \lesssim 10^7$ yr. {High values of $\delta$ may be found in
starburst environments, where high SN rates and possibly also top-heavy IMFs are expected. However, in general $\delta$ is likely small, since high rates of dust destruction
are somewhat inconsistent with the fact that dust is ubiquitous throughout the Universe.}

Although simplifying assumptions have been made in this study {in order to obtain a reasonably simple parametric model in terms of $\epsilon$ and $\delta$}, a clear
outwards slope is unlikely to be the result of any other mechanism than dust growth in the ISM. Other mechanisms, {which however} appear less effective:

\begin{itemize}
\item Accretion of dust free material onto the galactic disc may affect the dust-to-metals ratio if the infalling gas contains some fraction of atomic metals (see Appendix \ref{infall}
for further details and worked out examples). The metallicity of the accreted gas is likely much less than that of the ISM, so the effect cannot be very large and it would also mimic the effect of dust destruction rather 
than dust growth. 

\item Secondary dust production in stars, i.e., a stellar dust yield which increases as the metallicity of stars increases, may in principle create a dust-to-metals gradient along a galaxy disc. However, the relative increase 
of the stellar dust yield along the disc cannot be arbitrarily large. In particular, the dust-to-metals gradient can never become steeper than the metallicity gradient only owing to secondary dust production in stars 
(for further details and a more quantitative analysis, see Appendix \ref{secondarydust}). 

\item The lifetime of stars may also play a role, but since the very same stars that are producing the metals are also responsible for the stellar production of dust, this effect cannot be dominant. In fact, it should be
negligible.
\end{itemize}

Thus, we conclude that {\it dust-to-metals gradients can be used as a diagnostic for interstellar dust growth in galaxy discs, where a negative slope indicates dust growth}.\\

Dust growth has a non-linear nature as the time scale for it must depend on both the metallicity and the amount of available seed grains. As a consequence there is a "critical" metallicity (which depends
on the dust-growth and dust-destruction time scales as well as the dust-to-gas ratio) at which the dust production by interstellar grain growth exceeds stellar dust production and the dust-to-gas ratio diverges from 
the steady increase obtained in case the dust mass is owing to stars only. This allows for bends in the logarithmic slopes of the dust-to-metals profile even if the metallicity follows an exponential fall-off with 
galactocentric distance. Dust destruction in the ISM due to SNe may also affect the shape of the dust-to-metals profile, creating a central depression as the dust-to-gas ratio, the metallicity and the integrated number
of SNe typically increases in the central parts of a galaxy disc compared to the outer disc. However, since dust growth increases as well, the expected net effect is an increased dust-to-metals ratio in any case.

Finally, we note that combining recent observational results \citep{Munoz-Mateos09,Moustakas10} one finds that dust-to-metals gradients in late-type galaxy discs appear relatively steep (and negative), i.e., show 
a clear fall-off with galactocentric distance, which suggest interstellar dust growth is more important than stellar dust production.  In Paper II of this series, where we compare theoretical models and 
observational results in more detail, we return to this fact and look for more quantitative evidence of interstellar dust growth being the dominant dust production mechanism in late-type galaxies.

\section*{Acknowledgments}
The authors thank the reviewer, Anthony Jones, for constructive and helpful comments and criticism that greatly helped to improve the presentation.
L.M. acknowledges support from the Swedish Research Council (Vetenskapsr\aa det). The Dark Cosmology Centre is funded by the Danish National Research Foundation.

\appendix
\section{Effects of secondary stellar dust production}
\label{secondarydust}
It is important to remember that interstellar dust growth is not the only mechanism that can give rise to a dust-to-metals gradient. Secondary dust production in stars, i.e., a metallicity-dependent yield, could in principle
have similar effects. Splitting the stellar dust yield into two components, the constant primary yield $y_{\rm d}^{\rm p}$ and the metallicity-dependent secondary yield $y_{\rm d}^{\rm s}= y_{\rm d,\, \odot}^{\rm s}Z/Z_\odot$, 
and assuming there is no growth, nor destruction of dust in the ISM, we obtain
\begin{equation}
{dZ_{\rm d}\over dZ} = {1\over y_Z}\left[y_{\rm d}^{\rm p} + y_{\rm d,\,\odot}^{\rm s}{Z\over Z_\odot} \right],
\end{equation}
which, with the initial condition $Z_{\rm d}(0)=0$, has the solution 
\begin{equation}
Z_{\rm d} = {Z\over y_Z}\left(y_{\rm d}^{\rm p} + {y_{\rm d,\,\odot}^{\rm s}\over 2}{Z\over Z_\odot}\right).
\end{equation}
Using the notation introduced in section \ref{constraints}, we can also write
\begin{equation}
\Delta_\zeta = \left(1+2{y_{\rm d}^{\rm p}\over y_{\rm d,\,\odot}^{\rm s}}{1\over Z} \right)^{-1}\Delta_Z.
\end{equation}
In the limit where $y_{\rm d}^{\rm p} \ll y_{\rm d,\,\odot}^{\rm s} Z$ we then have $\Delta_\zeta \approx \Delta_Z$, which is the steepest dust-to-metals gradient obtainable for a given metallicity gradient. Hence, if the dust-to-metals
gradient is steeper than the metallicity gradient, then there must be dust growth in the ISM to account for that steepness. 

It is quite unlikely that $y_{\rm d}^{\rm p} \ll y_{\rm d,\,\odot}^{\rm s} Z$ as dust production in stars is likely primary to almost the same extent as the metals production is. Since most of the metals are primary, the secondary
dust component cannot be dominant, which implies $\Delta_\zeta < \Delta_Z$. In fact, modelling of stellar dust production suggest dust yields can be relatively high even at $Z=0$ \citep[see][and references therein]{Gall11b}.
It is actually reasonable to assume the secondary yield is no more (likely less) than 50\% of the primary yield at solar metallicity $Z_\odot$. If $y_{\rm d,\,\odot}^{\rm s} = y_{\rm d}^{\rm p}/2$, then 
\begin{equation}
\Delta_\zeta = \left(1+{4\over Z} \right)^{-1}\Delta_Z,
\end{equation}
where we note that $4/Z_\odot \approx 300$. More precisely, this implies $\Delta_\zeta$ is at least about two orders of magnitude smaller than $\Delta_Z$ for all reasonable metallicities along a galaxy disc. Thus, we conclude
that although the dust-to-metals gradient $\Delta_\zeta$ is technically non-zero in this case, it is still consistent with a flat dust-to-metals profile as metallicity gradients are rarely very steep \citep{Pilyugin04}. Secondary dust
production in stars cannot be responsible for a significant dust-to-metals gradient.

\section{Effects of infall}
\label{infall}
Throughout this paper we have treated the dust evolution in late-type galaxies assuming they are "closed boxes", i.e., that there is neither any inflow, nor any outflow of gas and metals to/from the disc. In reality
accretion of gas and minor mergers with smaller galaxies are important for the chemical evolution of a galaxy disc and thus also important for the shaping of the dust component. Hence, an infall component in
equation (\ref{dustz}) would have been in its place, but we omitted it for simplicity. However, as we will show here, the effect of infall is not such that it can qualitatively change any of our results. In fact, a theorem similar to that
presented in section \ref{constraints} could likely be formulated, but it would be less transparent as regarding the effects of growth and destruction of dust in the ISM.

In case of no growth or destruction of dust in the ISM and accretion of pristine (unenriched) gas at a rate $ \dot{\Sigma}_{\rm inf}$, which contains no metals in any form, we have the equation 
\begin{equation}
\label{dustzinf}
{dZ_{\rm d}\over dZ} = {y_{\rm d}-Z_{\rm d} A\over y_Z-Z\,A}, \quad A(r,t) \equiv \dot{\Sigma}_{\rm inf}(r,t)\left({d\Sigma_{\rm s}\over dt}\right)^{-1}.
\end{equation}
If $A$ is constant, the solution to the equation above is
\begin{equation}
Z_{\rm d} = {y_{\rm d}\over y_Z}\,Z,
\end{equation}
given the initial condition $Z_{\rm d}(0)=Z(0)=0$.
Hence, pristine infall likely does not affect the dust-to-metals ratio much. 

If the accreted gas contains metals the situation is quite different. Including metal-enriched infall, equation (\ref{dustzinf}) becomes
\begin{equation}
{dZ_{\rm d}\over dZ} = {y_{\rm d}-Z_{\rm d}A\over y_Z-(Z-Z_{\rm inf})\,A}.
\end{equation}
Assuming an outflow of interstellar gas where some of the metals in that gas are accreted back onto the disc, we can write $Z_{\rm inf} = \nu\,Z$ 
\citep[usually referred to as a "galactic fountain" model, see][]{Recchi08} the solution is
\begin{equation}
Z_{\rm d} = {y_{\rm d}\over A} \left\{1-\left[1+(\nu-1)\,A{Z\over y'_Z} \right]^{1\over 1-\nu} \right\},
\end{equation}
where $y'_Z$ is a reduced metal yield to account for the metals lost in the outflow.
For the specific case $\nu = 1/2$, we obtain the solution
\begin{equation}
{Z_{\rm d}\over Z} = {y_{\rm d}\over y'_Z}\left(1 - {A\over 4} {Z\over y'_Z}\right),
\end{equation}
from which it is easy to see that the effect of infall is reminiscent of the effect of dust destruction in the ISM due to SNe. i.e., that the dust-to-metals ratio decreases with metallicity (cf. equation \ref{nogrow}). 
Moreover, if the dust destruction term is included in equation (\ref{dustzinf}), we have
\begin{equation}
Z_{\rm d} = {y_{\rm d}\over A+\delta} \left\{1-\left[1+(\nu-1)\,A{Z\over y'_Z} \right]^{\delta\over 1-\nu} \right\},
\end{equation}
which is a solution of the same mathematical form as above. Hence, it is quite clear that metal-enriched infall has an effect which is very similar to that of dust destruction by SNe, which means that 
infall alone cannot create a dust-to-metals gradient with the same sign as the metallicity gradient.

\section{General solution of equation (34)}
\label{hypergeo}
The general solution to equation (\ref{dustz}) presented in section \ref{models} is expressed in terms of a product of the confluent hypergeometric Kummer-Tricomi functions $M$ and $U$ \citep{Kummer1837,Tricomi47}. 
This solution exist because equation (\ref{dustz}) is related to Kummer's equation (also known as the confluent hypergeometric equation), i.e.,
\begin{equation}
\label{Kummer}
z\frac{d^2w}{dz^2} + (b-z)\frac{dw}{dz} - aw = 0,
\end{equation}
which has the solution $w(z) = c_1 M(a,b;z)+ c_2 U(a,b;z)$, where $c_1$ and $c_2$ are arbitrary constants. With the variable change
\begin{equation}
\label{xidef}
\xi \equiv {1\over 2}{(\epsilon Z - \delta)^2\over \epsilon y_Z},
\end{equation}
equation (\ref{dustz}) can be rewritten as an equation of the form
\begin{equation}
{dZ_{\rm d}\over d\xi} = {y_{\rm d}\over\sqrt{\epsilon y_Z\xi}} + Z_{\rm d} - {\epsilon\over\sqrt{\epsilon y_Z\xi}}Z_{\rm d}^2.
\end{equation}
This is a Riccati equation, which has the general form
\begin{equation}
\label{Riccati}
{dy\over dx} = q_0(x) + q_1(x) \, y(x) + q_2(x) \, y^2(x).
\end{equation}
Such non-linear equations can be reduced to a second order linear ordinary differential equation \citep{Ince56} of the form
\begin{equation}
\label{Riccati2}
{d^2u\over dx^2}-R(x)\,{du\over dx} +S(x)\,u(x)=0,
\end{equation}
where
\begin{equation}
R(x) = q_1(x) + {1\over q_2(x)}{dq_2\over dx}, \quad S(x) = q_0(x)\,q_2(x).
\end{equation}
A solution to equation (\ref{Riccati2}) provides a solution to equation (\ref{Riccati}) as
\begin{equation}
y(x) = - {1\over q_2(x)\,u(x)}{du\over dx}.
\end{equation}
Identifying the Riccati coefficients as
\begin{equation}
q_0(\xi) = {y_d\over\sqrt{2\epsilon y_Z\xi}},\quad q_1(\xi) = 1, \quad q_2(\xi) = -{1\over\sqrt{2\epsilon y_Z \xi}},
\end{equation}
we find the associated second order linear ordinary differential equation,
\begin{equation}
\label{kummereq}
\xi{d^2u\over d\xi^2} + \left({1\over 2}-\xi\right){du\over d\xi} - {1\over 2}{y_{\rm d}\over y_Z}\,u(\xi) = 0.
\end{equation}
This is the Kummer equation for $b = 1/2$ and $a = y_{\rm d}/2y_Z$. Reverse Riccati reduction and back-substitution, together with the natural initial condition $Z_{\rm d}(0)=0$, will provide the general solution
given in section \ref{models} after some algebra.

The Kummer equation has an awkward property: it has a regular (order one) singularity at the origin (at $\xi = 0$ in the case above). This means that no solution exists at $\xi=0$ (or $Z = \delta/\epsilon$) and that 
the region near this point must be avoided when applying this solution. In particular, when the Kummer-Tricomi functions are implemented numerically, the algorithm for computing them will be unstable in the 
vicinity of this singular point. We have therefore considered  the special (and not entirely realistic) case where stellar dust  production is negligible from a point in time when the metallicity has a certain value $Z_0$
and the dust-to-metals  ratio has a value $\zeta_0$. Assuming $y_{\rm d}\to 0$ equation (\ref{dustz}) reduces to
\begin{equation}
y_Z\,{dZ_{\rm d}\over dZ}  = .(\epsilon Z-\delta)\,Z_{\rm d}-\epsilon Z_{\rm d}^2.
\end{equation}
Riccati reduction as above yields
\begin{equation}
{d^2u\over d\xi^2}-\left({1\over 2 - \xi}\right)\,{du\over d\xi} =0,
\end{equation}
where $\xi$ is as previously defined (equation \ref{xidef}). This equation is non-singular, but requires that $Z>Z_0$, where $Z_0$ is some finite initial value. The general solution is
\begin{equation}
u(\xi) = C_0 + C_1 \sqrt{\pi}\, {\rm erfi}\left(\sqrt{\xi}\right),
\end{equation}
where $C_0$, $C_1$ are constants to be fixed by initial conditions as we do reverse the Riccati reduction and back-substitute. With $Z_0 = Z(t_0) \neq 0$ and $Z_{{\rm d},\,0} = Z_{\rm d}(0) \neq 0$ as the initial 
conditions it is then possible to obtain the solution given as equation (\ref{nostars}) in section (\ref{specialcases}).

\section{Numerical implementation of the Kummer-Tricomi functions}
\label{numerical}
The Kummer-Tricomi functions, used in section \ref{models} and Appendix \ref{hypergeo} above, can be defined in terms of integral quantities \citep{Kummer1837,Tricomi47},
\begin{equation}
M(a,b;z)= \frac{\Gamma(b)}{\Gamma(a)\Gamma(b-a)}\int_0^1 e^{zu}u^{a-1}(1-u)^{b-a-1}\,du,
\end{equation}
for $\Re(b) > \Re(a) > 0$,
\begin{equation}
U(a,b;z) = \frac{1}{\Gamma(a)}\int_0^\infty e^{-zt}t^{a-1}(1+t)^{b-a-1}\,dt,
\end{equation}
for $\Re(a)>0$. This is not very convenient for numerical implementation though. As alternative we can consider the following.
The function $M$ is identical to the ${}_1F_1$ hypergeometric function which can be defined as an infinite series,
\begin{equation}
M(a,b;z) = \sum_{n=0}^\infty \frac {a^{(n)} z^n} {b^{(n)} n!}={}_1F_1(a,b;z)
\end{equation}
where
\begin{equation}
a^{(n)}=a(a+1)(a+2)\cdots(a+n-1) =\frac{\Gamma(a+n)}{\Gamma(a)},
\end{equation}
is the Pochhammer symbol and $\Gamma$ is the Gamma function,
\begin{equation}
 \Gamma(z) = \int_0^\infty  t^{z-1} e^{-t}\,dt.
 \end{equation}
The function $M$ can thus be implemented numerically by computing the above series until some arbitrary precision is obtained. The typical number of terms needed to reach the 
precision limit of a standard Intel processor is at most a few hundred. This may still cause problems when computing the Pochhammer symbol, since this will have to be done using some limited implementation 
of $\Gamma$ to obtain reasonable computation speed. The basic issue is the fact that $\Gamma$, as well as the factorial, is usually not implemented for large arguments. For example, in IDL and MATLAB the 
argument $z$ cannot exceed $\sim 170$. However, this situation rarely occurs.

The function $U$ can be defined in terms of the function $M$ \citep{Tricomi47} by
\begin{eqnarray}
U(a,b;z)=\frac{\Gamma(1-b)}{\Gamma(a-b+1)}M(a,b;z)+\\\nonumber
\frac{\Gamma(b-1)}{\Gamma(a)}z^{1-b}M(a-b+1,2-b;z), 
\end{eqnarray}
which is straight forward to implement, except for integer $b$ (where $U$ is not defined). The general solution to equation (\ref{dustz}) does not give rise to any integer values for the $b$, so we will not consider how 
to implement the analytical extension of $U$ for integer $b$. This can be done, but goes beyond the scope of this study.

\end{document}